\documentclass[12pt]{article}
\usepackage{graphicx,amsmath,amssymb,url,enumerate,mathrsfs,epsfig,color}
\usepackage{tikz}
\usepackage{float}
\usetikzlibrary{lindenmayersystems}

\usepackage{amsmath}
\usepackage{amssymb}
\usepackage{yfonts}

\usepackage{tikz,pgfplots}

\usetikzlibrary{calc, patterns,arrows.meta, arrows, shapes.geometric}

\usepackage{graphicx,amsmath,amssymb,url,enumerate,mathrsfs,epsfig,color}

\usetikzlibrary{decorations.text}
\usetikzlibrary{decorations.markings}

\pgfplotsset{compat=1.8}

\usepackage{enumerate}
\usepackage{boxedminipage}
\usepackage{makeidx}
\usepackage{multicol}
\usepackage{xcolor}
\usepackage{mathtools}
\usepackage{comment}

\usetikzlibrary{calc, patterns,arrows, shapes.geometric}
\def\QED{\hskip0.1em\hfill\null\ \null\nobreak\hfill
\kern3pt\lower1.8pt\vbox{\hrule\hbox
{\vrule\kern1pt\vbox{\kern1.7pt \hbox{$\scriptstyle
QED$}\kern0.2pt}\kern1pt\vrule}\hrule}}
\newtheorem{rem}{Remark}[section]

\title{On a generalized principle of fractal stiffness self-similarity}
\author{Marcelo Epstein\footnote{University of Calgary, Calgary, Canada \newline email: mepstein@ucalgary.ca}  }
\date{}
\begin{document}
\maketitle

\begin{abstract} The principle of fractal stiffness self-similarity is expanded to encompass structures with several differently-scaled contributors to the total stiffness matrix. The generalized principle is applied to solve the problem of a fractal triangular gasket that incorporates drilling modes, with a view to further applications to the modelling of fractal shells.

\end{abstract}
\bigskip


\section{Introduction}

Whether fractality has a stronger or weaker claim than continuity for the faithful representation of real-life structures is a question whose answer is better left to philosophers of science. The much more modest aim of this article is to offer a concrete numerical answer to the problem of finding an exact expression of the in-plane stiffness matrix of a material Sierpi\'nski triangle endowed with linear material properties in the small deformation regime, when drilling degrees of freedom are taken into consideration. The 9 degrees of freedom with respect to which the stiffness is evaluated are the in-plane displacements and rotations of the 3 vertices of the triangle. The adjective `exact' is purposely used to emphasize in which two senses the theory presented is not an approximation. In the first place, since the Sierpi\'nski triangle \cite{falconer} is a self-similar object with an infinite number of hierarchies of self-replication, an approximation might have limited itself to the consideration of a finite number of levels in the hierarchy upon which a classical elastic analysis is carried out. The second sense in which the derivation herein is not an approximation is that the triangle is not in any way to be interpreted as some kind of finite element whose systematic decrease in size might lead to convergence to some ultimate reality. On the contrary, the triangle is a structure in an of itself which, if so desired, may be connected at its vertices with other similar or different structures.

Among more fundamental studies of fractals as load-bearing media we may mention the works of Carpinteri et al. \cite{carp1} and Carpinteri and Cornetti \cite{carp2} (based upon fractional calculus \cite{mainardi, oldham}), Tarasov \cite{tarasov} and Ostoja-Starzewski \cite{ostoja1} (using a generalized integral calculus on fractals), and Epstein and \'Sniatycki \cite{eands} (attempting to apply Sikorski's  theory of differential spaces). The down-to-earth structural approach presented initially in \cite{paper1}, based on considerations of symmetry, equilibrium, scaling, and self-similarity alone, was applied in \cite{plate} to derive the stiffness of a triangular fractal plate under transversal loading. To make it possible to eventually extend the formulation to a fully fledged shell element the in-plane degrees of freedom need to be incorporated, including the all-important drilling modes ensuring that moment balance can be satisfied at the meeting points between triangles. Section \ref{sec:simplest} is entirely devoted to demonstrate the somewhat unexpected consequences of self-similarity when applied to the elementary example of a classical beam. In this simple situation, the classical stiffness matrix is obtained by adopting a principle of stiffness self-similarity, introduced in \cite{paper1} in an application to some elementary fractal structures. Section \ref{sec:frame} is devoted to show how a triangular frame (not a fractal), in which rotational degrees of freedom are taken into consideration, exhibits two different stiffness scaling patterns, associated, respectively, with stretching and bending. This example is used as a motivating springboard to deal in Section \ref{sec:gasket} with more complex fractal structures, such as when drilling degrees of freedom are introduced at the nodes of a Sierpi\'nski gasket. This example provides us with an occasion of introducing a generalized principle of multi-scaled stiffness self-similarity. The application of this principle to the Sierpi\'nski gasket is described in detail and the final expression of the total stiffness matrix is obtained. Some final thoughts about the features and limitations of this approach are the subject of Section \ref{sec:final}.

\section{The simplest example}
\label{sec:simplest}

\subsection{The classical beam}

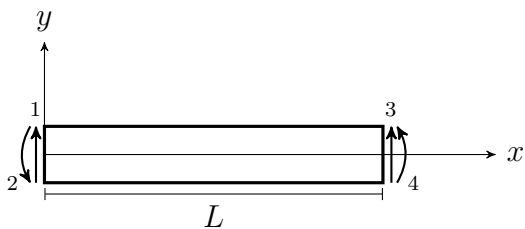
\begin{figure} [H]
\begin{center}
\begin{tikzpicture} [scale=0.75]
\draw[-stealth'] (0,0)--(8,0);
\node[right] at (8,0){$x$};
\draw[-stealth'] (0,0)--(0,2);
\node[above] at (0,2) {$y$};
\draw[very thick] (0,-0.5)--(6,-0.5)--(6,0.5)--(0,0.5)--cycle;
\draw [thick, -stealth'] (-0.15,-0.5)--(-0.15,0.5);
\node[above] at (-0.15,0.5) {$_1$};
\draw [thick, -stealth'] (6.15,-0.5)--(6.15,0.5);
\node[above] at (6.15,0.5) {$_3$};
\draw [thick, stealth'-] (-0.25,-0.5) to [bend left] (-0.25,0.5);
\node[left] at (-0.25,-0.5) {$_2$};
\draw [thick, -stealth'] (6.25,-0.5)to [bend right] (6.25,0.5);
\node[right] at (6.25,-0.5) {$_4$};
\draw[|-|] (0,-0.7)--(6,-0.7);
\node[below] at (3,-0.7) {$L$};
\end{tikzpicture}
\end{center}
 \caption{An Euler-Bernoulli beam}
\label{fig:euler}
\end{figure}

Let us consider the classical Euler-Bernoulli beam and its stiffness well-known matrix $[K]_e$ in the form
\begin{equation} \label{eq1}
[K]_e=EI \left[
\begin{matrix}
12/ L^3& 6 /L^2&-12/L^3 &6/L^2\\6/L^2&4/L&-6/ L^2&2/ L \\ -12/ L^3&-6/ L^2&12/ L^3&-6/ L^2\\6/ L^2&2/ L&-6/ L^2&4/L
\end{matrix}
\right]
\end{equation}
In this equation, only the 4 nodal degrees of freedom (two translations and two rotations) shown in Figure \ref{fig:euler} have been considered. The beam cross section is constant and symmetric with respect to the plane of the figure. The moment of inertia of the cross section with respect to the centroidal axis perpendicular to this plane is denoted by $I$. The material abides by Hooke's law with elastic modulus $E$. The beam deflections are assumed to be very small when compared with the beam length $L$. Moreover, the assumption is made that the cross sections remain perpendicular to the deflected axis.
Because of the inhomogeneous nature of the physical units of displacements and rotations (and their static counterparts, forces and torques) it is convenient to introduce a diagonal {\it scaling matrix} $\Lambda$
\begin{equation}\label{eq1a}
[\Lambda]=\left[
\begin{matrix}
1&0&0&0\\0&L&0&0\\0&0&1&0\\0&0&0&L
\end{matrix}
\right]
\end{equation}
and the non-dimensional stiffness matrix
\begin{equation}
[\kappa]_e= \left[
\begin{matrix}
12& 6 &-12&6\\6&4&-6&2 \\ -12&-6&12&-6\\6&2&-6&4
\end{matrix}
\right].
\end{equation}
in whose terms the stiffness matrix can be recast as
\begin{equation}
[K]_e=\frac{EI}{L^3}\;[\Lambda] [\kappa]_e [\Lambda]
\end{equation}
The matrix $[\kappa]_e$ is invariant under changes in length.

\subsection{Symmetry and statics}

The specific form (\ref{eq1}) of the stiffness matrix $[K]_e$ is usually obtained by a careful implementation of various kinematic and constitutive assumptions followed by the integration of a differential equation of equilibrium. Could it have been derived otherwise?
We cannot help but notice that our beam, by virtue of the constancy of its cross section and the uniformity of its material properties, enjoys some a priori symmetry properties, regardless of the specific assumptions embodied by the Euler-Bernoulli theory. As a consequence of these geometric and material symmetries alone, we could have predicted that the (symmetric) stiffness matrix should be of the form
\begin{equation} \label{eq2}
[K]=\left[
\begin{matrix}
a&b&d&e\\b&c&f&g\\d&f&a&-b\\e&g&-b&c
\end{matrix}
\right]
\end{equation}
where $a,b,c,d,e,f,g$ are constants. Moreover, by the very meaning of the stiffness coefficients (reactions in the restrained structure due to unit value of one of the degrees of freedom) each of the columns of $[K]$ represents a system of forces and couples in equilibrium. These conditions can be represented compactly as
\begin{equation} \label{eq3}
[S]_L [K]=[0],
\end{equation}
where the $2\times 4$ static matrix $[S]_L$ has been chosen as
\begin{equation}
[S]_L= \left[
\begin{matrix}
1&0&1&0\\0&1&L&1
\end{matrix}
\right]
\end{equation}
As a direct consequence of these equilibrium considerations, the stiffness matrix can be reduced to
\begin{equation} \label{eq5}
[K]=\left[
\begin{matrix}
a&aL/2&-a&aL/2\\aL/2&c&-aL/2&aL^2/2 - c\\-a&-aL/2&a&-aL/2\\aL/2&aL^2/2 - c&-aL/2&c
\end{matrix}
\right]
\end{equation}
In short, on the basis of symmetry and statics alone, we have reduced the stiffness matrix to a dependence on only two material constants, $a$ and $c$. Introducing the non-dimensional ratio
\begin{equation} \label{eq9}
\gamma=\frac{c}{aL^2},
\end{equation}
we can write
\begin{equation}
[K]= a [\Lambda] [\kappa] [\Lambda],
\end{equation}
where $[\kappa]$ is the non-dimensional matrix
\begin{equation}
[\kappa]=\left[
\begin{matrix}
1&1/2&-1&1/2 \\ 1/2&\gamma&-1/2&1/2-\gamma\\-1&-1/2&1&-1/2\\1/2&1/2-\gamma&-1/2&\gamma
\end{matrix}
\right]
\end{equation}
We observe that the matrix $[K]_e$ in Equation (\ref{eq1}) depends on a single material constant, whereas our $[K]$ is governed by two constants. To reduce these two constants to a single one we proceed to exploit the property of {\it self-similarity}.

\subsection{Geometric self-similarity}
\label{sec:condense}

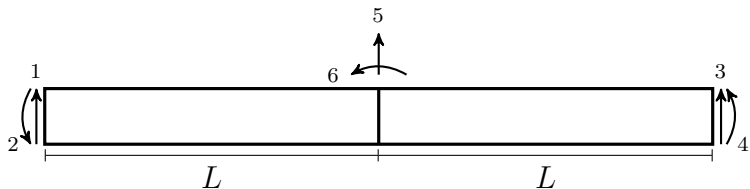
\begin{figure} [H]
\begin{center}
\begin{tikzpicture} [scale=0.74]
\draw[very thick] (0,-0.5)--(12,-0.5)--(12,0.5)--(0,0.5)--cycle;
\draw[very thick] (6,-0.5)--(6,0.5);
\draw [thick, -stealth'] (-0.15,-0.5)--(-0.15,0.5);
\node[above] at (-0.15,0.5) {$_1$};
\draw [thick, -stealth'] (12.15,-0.5)--(12.15,0.5);
\node[above] at (12.15,0.5) {$_3$};
\draw [thick, stealth'-] (-0.25,-0.5) to [bend left] (-0.25,0.5);
\node[left] at (-0.25,-0.5) {$_2$};
\draw [thick, -stealth'] (12.25,-0.5)to [bend right] (12.25,0.5);
\node[right] at (12.25,-0.5) {$_4$};
\draw[|-|] (0,-0.7)--(6,-0.7);
\node[below] at (3,-0.7) {$L$};
\draw[-|] (6,-0.7)--(12,-0.7);
\node[below] at (9,-0.7) {$L$};
\draw [thick, -stealth'] (6,0.75)--(6,1.5);
\draw [thick, -stealth'] (6.5,0.75)to [bend right] (5.5,0.75);
\node[above] at (6,1.5) {$_5$};
\node[left] at (5.5,0.75) {$_6$};
\end{tikzpicture}
\end{center}
 \caption{Self-similarity}
\label{fig:condense}
\end{figure}

Assuming that the constants $a$ and $c$ have been determined (experimentally, say), can we infer anything at all for a beam identical to the one under consideration except that its length has been doubled? Clearly, its stiffness matrix, $[\hat K]$ say, will also be of the form (\ref{eq5}) with some new coefficients, $\hat a$ and $\hat c$. Are these `hatted' coefficients somehow related to their non-hatted counterparts? To answer this question we may try to resort to the fact that our beam enjoys the property of {\it self-similarity}. Indeed, the beam of length $2L$ can be obtained by gluing together end-to-end two identical beams of length $L$, At the common point, the translational and rotation degrees of freedom are shared, as shown in Figure \ref{fig:condense}, once continuity is enforced. These two internal degrees of freedom can then be eliminated by a process of {\it static condensation} \cite{guyan}.
For convenience, let us partition the stiffness matrix $[K]$ into four $2 \times 2$ matrix blocks, namely,
\begin{equation}
[K]=\left[
\begin{array} {cc}
A&B  \\  B^T&C
\end{array}
\right]
\end{equation}
Blocks $A$ and $C$ represent, respectively, the interactions of degrees of freedom $1,2$ and $3,4$, while block $B$ contains the mutual interactions between the first and the second pair. The stiffness matrix $[K']$ of the assembled 6-degree-of-freedom  structure in Figure \ref{fig:condense} is, therefore, obtained in block form as
\begin{equation}
[K']=\left[
\begin{array} {ccc}
A&0&B  \\ 0&C&B^T\\B^T& B&C+A
\end{array}
\right] = \left[
\begin{array} {cc}
A'&B'  \\ B'^T&C'
\end{array}
\right],
\end{equation}
where $A'$ is a $4 \times 4$ block and $C'$ is a $2 \times 2$ block. The extra degrees of freedom ($5$ and $6$) can be eliminated algebraically to deliver the $4 \times 4$ stiffness matrix
\begin{equation} 
[\hat K]=[A']-[B'] [C']^{-1}[B']^T=\left[
\begin{array} {cc}
\hat A&\hat B  \\  \hat B^T&\hat C
\end{array}
\right].
\end{equation}
Carrying out the operations indicated we obtain (for our example )
\begin{equation} \label{eq15}
[\hat K]= \frac{4c-aL^2}{8c}\left[
\begin{matrix}
a&aL&-a&aL \\aL&c+aL^2&-aL&-c+aL^2\\-a&-aL&a&-aL\\aL&-c+aL^2&-aL&c+aL^2
\end{matrix}
\right]
\end{equation}
We observe that the entries in this matrix satisfy both the symmetry conditions (\ref{eq2}) and the equilibrium conditions (\ref{eq3}) using the static matrix $[S]_{2L}$.

\subsection{A leap of faith}

It is at this point that we are called to make an epistemological leap of faith. By our considerations of symmetry and equilibrium for a beam of length $2L$ there exist two constants, $\bar a$ and $\bar c$, such that the stiffness matrix is expressible as
\begin{equation} \label{eq16}
[\bar K]=\left[
\begin{matrix}
\bar a&\bar a L&-\bar a&\bar a L\\\bar a L&\bar  c&-\bar a L&2 \bar a L^2 - \bar c\\-\bar a&-\bar a L&\bar  a&-\bar a L\\\bar a L&2 \bar a L^2 -\bar c&- \bar a L&\bar c
\end{matrix}
\right]
\end{equation}
This matrix can expressed in terms of a non-dimensional counterpart $[\bar \kappa]$ as
\begin{equation}
[\bar K]=\bar a [\bar \Lambda ] [\bar \kappa] [\bar \Lambda],
\end{equation}
with
\begin{equation}
[\bar \kappa]= \left[
\begin{matrix}
1&1/2&-1&1/2\\1/2& \bar \gamma&-1/2&1/2- \bar \gamma\\-1&-1/2&1&-1/2\\1/2&1/2-\bar \gamma&-1/2&  \bar \gamma
\end{matrix}
\right]
\end{equation}
where $[\bar \Lambda]$ and $\bar \gamma$ are the counterparts of (\ref{eq1a}) and (\ref{eq9}) when $L$ is replaced with $2L$, and $a$ and $c$ are replaced, respectively, with $\bar a$ and $\bar c$. As expected, $[\bar \kappa]$ is of the same form as $[\kappa]$. In the absence of any notion of absolute scale, we may assume that the non-dimensional matrix $[\bar k]$ is the same as $[\kappa]$. In other words, we assume that 
\begin{equation} \label{eq19}
\bar \gamma = \gamma,
\end{equation}
and consequently
\begin{equation}
\frac{\bar c}{\bar a}=4\frac{c}{a} 
\end{equation}
But since the stiffness of the double-length beam is represented both by $[\hat K]$ in Equation (\ref{eq15}) and by $[\bar K]$ in Equation (\ref{eq16}), we must conclude that
\begin{equation}
\frac{c}{a}=\frac{L^2}{3}
\end{equation}
and
\begin{equation}
\bar a=\frac{a}{8}
\end{equation}
It seems appropriate to call the ratio ${\bar a}/a$ the {\it stiffness scaling} ratio. As a result of these considerations, we obtain that the stiffness matrix of a bar of length $L$ must of necessity be of the form
\begin{equation}
[K]=a \left[
\begin{matrix}
1&L/2&-1&L/2\\L/2&L^2/3&-L/2&L^2/6\\-1&-L/2&1&-L/2\\L/2&L^2/6&-L/2&L^2/3
\end{matrix}
\right]
\end{equation}
The classical Euler beam stiffness matrix is recovered by setting $a=EI/L^3$. It is noteworthy that, beyond the choice of nodal degrees of freedom, no kinematic treatment has been invoked, but just considerations of symmetry, equilibrium, and self-similarity. The crucial assumption embodied in Equation (\ref{eq19}) can be regarded as a manifestation of a {\it principle of stiffness self-similarity} \cite{paper1}. It states that a change of geometrical scale brings about a concomitant change in the scale of the stiffness matrix. In our example, a doubling of the beam length results in a proportionate weakening of the stiffness by a factor of 8. 

\subsection{Remarks on stiffness self-similarity}

The fact that the stiffness matrix of a self-similar structure seems to emerge out of thin air is particularly noteworthy in the case of structures of a fractal nature. Indeed, the description of the displacement field within a fractal is a difficult mathematical problem. It cannot be approximated meaningfully by smooth functions, which would allow for a rational deduction of the stiffness properties via a constitutive law. The simple beam example just presented seems to suggest that, at least for self-similar geometries, it may be possible to avoid any explicit reference to the displacement field by invoking the principle of stiffness self-similarity. This approach was introduced in \cite{paper1} and exploited to obtain explicit expressions for the stiffness matrix of some elementary self-similar fractals ,such as the Sierpi\'nski gasket with 6 degrees of freedom (two displacement components at each vertex of the triangle).

It should be clear that, had we included in our simple beam example the axial deformations, there would have been two different elastic mechanisms at play, each one governed by a different geometric property (the cross-sectional moment of inertia and the area, respectively, for bending and axial effects). Similarly, if we had considered in the Sierpi\'nski gasket not just degrees of freedom of displacement but also in-plane rotations, the stiffness self-similarity adduced by the original principle would have been lost.

\begin{comment}
For the case of a Sierpi\'nski gasket and other simple fractals, the stiffness matrix has been derived \cite{paper1} from the principle of stiffness self-similarity, following the same procedure illustrated above for the case of a beam. A recent article \cite{recent} demonstrates the inapplicability of the original version of the similarity principle for certain generalizations of the von Koch beam.
\end{comment}

In the next section we will demonstrate, with a simple example of a non-fractal nature, the separate effects of two different mechanisms contributing to the total stiffness of a linear structure, each one with its own stiffness scaling property. This idea will then be used as a basis for the generalization of the principle of stiffness self-similarity and its application to an important example of a deformable fractal with drilling-mode degrees of freedom.

\section{Multi-scaling in a non-fractal structure}
\label{sec:frame}

\subsection{A frame}

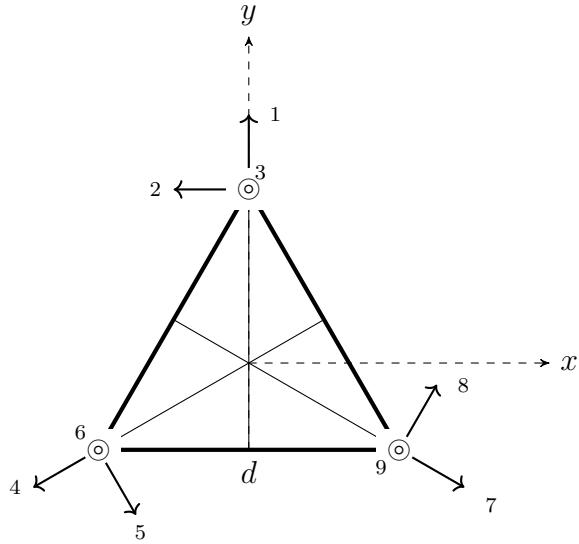
\begin{figure} [H]
\begin{center}
\begin{tikzpicture}
\node at (0,0) {$\odot$};
\node at (4,0) {$\odot$};
\node at (2,{2*sqrt(3)}) {$\odot$};
\draw[->,thick] (2,{0.2+2*sqrt(3)})--(2,{1+2*sqrt(3)});
\draw[->,thick] (1.9,{2*sqrt(3)})--(1,{2*sqrt(3)});
\draw[dashed, -stealth'] (2.0,{2*sqrt(3)/3})--(6,{2*sqrt(3)/3});
\draw[dashed, -stealth'] (2,0)--(2,5.5);
\node[right] at (6,{2*sqrt(3)/3}) {$x$};
\node[above] at (2,5.5) {$y$};

\draw[ultra thick] (0,0)--(4,0)--( (2,{2*sqrt(3)})--cycle;
\draw (2,0)--( 2,{2*sqrt(3)});
\draw (0,0)--( 3,{sqrt(3)});
\draw (4,0)--( 1,{sqrt(3)});
\node[fill=white] at (0,0) {$\circledcirc$};
\node[fill=white] at (4,0) {$\circledcirc$};
\node[fill=white] at (2,{2*sqrt(3)}) {$\circledcirc$};
\node[above right] at (1.8,{2*sqrt(3)}) {$~_{3}$};
\node[right] at (2,{1+2*sqrt(3)}) {$~_{1}$};
\node[left] at (1,{2*sqrt(3)}) {$~_{2}$};
\begin{scope} [rotate around={120:(0,0)}]
\draw[->,thick] (0,0.2)--(0,1);
\draw[->,thick] (-0.2,0)--(-1,0);
\node[left] at (0,1) {$~_{4}$};
\node[above left] at (0,0) {$~_{6}$};
\node[below] at (-1,0) {$~_{5}$};
\end{scope}
\begin{scope} [rotate around={240:(4,0)}]
\draw[->,thick] (4,0.2)--(4,1);
\draw[->,thick] (3.8,0)--(3,0);
\node[below right] at (4,1) {$~_{7}$};
\node[below left] at (4,0) {$~_{9}$};
\node[right] at (3.0,0) {$~_{8}$};

\end{scope}
\node[below] at (2,0) {$d$};
\end{tikzpicture}
\end{center}
 \caption{The equilateral frame with its 9 degrees of freedom}
\label{fig:frame}
\end{figure}

To illustrate the stiffness implications of the coexistence of two different elastic mechanisms we consider an equilateral triangular frame made of three identical classical beams, as shown in Figure \ref{fig:frame}. The degrees of freedom are the in-plane displacements and rotations of the nodes. This is an elementary problem in structural analysis, but we will tackle it with a view to further inferences. To this end, we align the degrees of freedom of each node with a local right-handed Cartesian coordinate system with one axis along the median at the vertex, as shown in the figure. The in-plane nodal rotation is represented with a circled circle to indicate that it is a rotation vector perpendicular to the plane of the drawing and pointing towards the reader. The degrees of freedom have been numbered sequentially, node by node.

\subsection{Symmetry considerations}
\label{sec:symmetry}

\begin{figure} [H]
\begin{center}
\begin{tikzpicture}
\foreach \x in {0,1,2,3}
{\draw (0,\x)--(3,\x);
\draw (\x,0)--(\x,3);}
\node at (0.5,2.5) {$A$};
\node at (1.5,2.5) {$B$};
\node at (4.5,1.5) {$\Longrightarrow$};
\begin{scope} [shift={(6,0)}]
\foreach \x in {0,1,2,3}
{\draw (0,\x)--(3,\x);
\draw (\x,0)--(\x,3);}
\node at (0.5,2.5) {$A$};
\node at (1.5,2.5) {$B$};
\node at (0.5,1.5) {$B^T$};
\end{scope}
\end{tikzpicture}
\end{center}
 \caption{The fundamental blocks}
\label{fig2}
\end{figure}
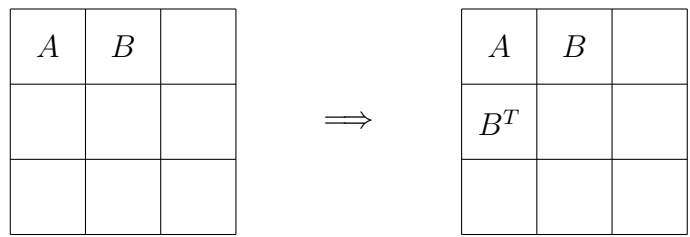

The $9 \times 9$ stiffness matrix $K$ of the frame will be partitioned into nine $3 \times 3$ blocks. In Figure \ref{fig2}, two of these blocks have been designated with the letters $A$ and $B$. Block $A$ represents the interactions between the 3 degrees of freedom of the uppermost node, while block $B$ contains the cross influences between these 3 degrees of freedom and those of the lower left node. Since the total stiffness matrix of an elastic structure is symmetric, so must be block $A$, that is $A^T=A$. The symmetry of $K$ also implies that the block just below $A$ must be $B^T$, as suggested on the right of Figure \ref{fig2}.

By the rotational symmetries of Figure \ref{fig:frame}, a symmetry that includes both the shape and the choice of nodal degrees of freedom. it is clear that the blocks along the main diagonal must all be equal to $A$.  Moreover, the relation between degrees of freedom $1,2,3$ and $4,5,6$ (described by block $B$) must be identical to their counterparts between $4,5,6$ and $7,8,9$. Finally, we observe that in Figure \ref{fig2} the block $B^T$ describes the influences between the degrees of freedom of a node and those of the node situated at $120^o$ clockwise from it. Since this is precisely the case between the uppermost node and the lower right node, we conclude that the corresponding block is $B^T$ as well. The complete stiffness matrix $K$ can now be fully populated as
\begin{equation} \label{eq:ab}
\begin{tikzpicture}
\node at  (-1,1.5) {$[K]\;=$};
\draw[thick] (-0.1,0.1)--(-0.1,2.9);
\draw[thick] (-0.1,0.1)--(0,0.1);
\draw[thick] (-0.1,2.9)--(0,2.9);
\draw[thick] (3.1,0.1)--(3.1,2.9);
\draw[thick] (3,0.1)--(3.1,0.1);
\draw[thick] (3,2.9)--(3.1,2.9);
\foreach \x in {1,2}
{\draw (0,\x)--(3,\x);
\draw (\x,0.2)--(\x,2.8);}
\node at (0.5,2.5) {$A$};
\node at (1.5,2.5) {$B$};
\node at (0.5,1.5) {$B^T$};
\node at (2.5,2.5) {$B^T$};
\node at (1.5,1.5) {$A$};
\node at (2.5,1.5) {$B$};
\node at (0.5,0.5) {$B$};
\node at (1.5,0.5) {$B^T$};
\node at (2.5,0.5) {$A$};
\end{tikzpicture}
\end{equation}

We can proceed to analyse the two fundamental blocks, $A$ and $B$. The symmetry of the triangle about the vertical line passing through the apex implies that the stiffness coefficients $k_{12}$ and $k_{13}$ vanish. Consequently, the matrix block $A$ has the form
\begin{equation} \label{eq25}
[A]=\left[
\begin{matrix}
a_1&0&0 \\0&a_2&a_4\\ 0&a_4&a_3
\end{matrix}
\right]
\end{equation}
For the same reason\footnote{The mere vanishing of $a_5$ and $a_6$ and the equilibrium conditions described in Section \ref{sec:equilibrium} are enough to arrive at the same conclusions.}, we must have $k_{24}=-k_{15}$, $k_{34}=-k_{16}$, and $k_{35}=k_{26}$, yielding
\begin{equation} \label{eq26}
[B]=\left[
\begin{matrix}
b_1&b_2&b_3 \\-b_2&b_4&b_5\\ -b_3&b_5&b_6
\end{matrix}
\right]
\end{equation}

\subsection{Equilibrium}
\label{sec:equilibrium}

If we recall that the physical meaning of the stiffness coefficient $k_{ij}$ is the reaction brought about in correspondence with the degree of freedom number $i$ when the degree of freedom number $j$ in the otherwise restrained structure is given a unit value (of displacement or rotation), we immediately realize that the entries in each column $j$ must satisfy the equilibrium conditions of a system of forces and couples. In our case, due to the nature of the reactions, these are three equations, namely, the vanishing of the sum of forces in two different direction in the plane, and the sum of moments with respect to an axis perpendicular to it.

The equilibrium equations of sum force components relative to the $x$ and $y$ axes and the sum of moments about the top vertex is enforced by multiplication of $[K]$ to the left by the following $3\times 9$ matrix
\begin{equation}
[S]=\left[
\begin{matrix}
0 & -1 & 0 & -\sqrt{3}/2 & 1/2 &0 &  \sqrt{3}/2  &1/2 &0 \\
1 & 0 & 0 & -1/2 & -\sqrt{3}/2 &0 &  -1/2 &  \sqrt{3}/2  & 0 \\
0 & 0 & 1& -d/2 &  \sqrt{3}d/2 & 1 & d/2 &  \sqrt{3}d/2 & 1
\end{matrix}
\right],
\end{equation}
where $d$ is the length of the triangle side. Thus, the equilibrium conditions can be expressed as the matrix equation
\begin{equation} \label{eq4}
[S][K]=[0],
\end{equation}
in which the right-hand side is the $3 \times 9$ zero matrix. Carrying out the matrix product we obtain 5 independent linear equations that can be written as
\begin{equation}
\begin{split}
a_1-b_1-\sqrt{3}b_2=0 \\
-a_2+\sqrt{3} b_2+b_4=0\\
-a_4+\sqrt{3} b_3+b_5= 0 \\
a_4+2 b_5+(b_2+\sqrt{3} b_4)d= 0 \\
a_3+2 b_6 +(b_3+\sqrt{3} b_5) d= 0
\end{split}
\end{equation}
We use these equations to eliminate $b_2, b_3, b_4, b_5$ and $b_6$ and express $[B]$ as
\begin{equation} \label{eq:equil}
\begin{split}
b_2&=\frac{\sqrt{3}}{3} (a_1-b_1) \\
b_3&=\frac{\sqrt{3}}{2}a_4 -\left(\frac{1}{3}(a_1-b_1) -\frac{1}{2} a_2\right)d\\
b_4&=a_2-(a_1-b_1)\\
b_5&=-\frac{1}{2}a_4+\sqrt{3}\left(\frac{1}{3}(a_1-b_1)-\frac{1}{2}a_2 \right)d\\
b_6&=-\frac{1}{2}a_3-\left(\frac{1}{3}(a_1-b_1)-\frac{1}{2}a_2\right)d^2
\end{split}
\end{equation}
In this way the stiffness matrix $[K]$ has been reduced to depend only on 5 independent coefficients, namely, $a_1, a_2,a_3, a_4$ and $b_1$. 

\subsection{The constitutive law}

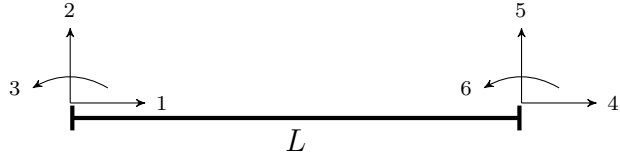
\begin{figure}[H]
\begin{center}
\begin{tikzpicture}
\draw[ultra thick, |-|] (0,0)--(6,0);
\draw[-stealth'] (0,0.2)--(1,0.2);
\node[right] at (1,0.2) {$_1$};
\draw[-stealth'] (0,0.2)--(0,1.2);
\node[above] at (0,1.2) {$_2$};
\draw[-stealth'] (0.5,0.4) to  [bend right] (-0.5,0.4);
\node[left] at (-0.5,0.4) {$_3$};
\begin{scope} [xshift=6cm]
\draw[-stealth'] (0,0.2)--(1,0.2);
\node[right] at (1,0.2) {$_4$};
\draw[-stealth'] (0,0.2)--(0,1.2);
\node[above] at (0,1.2) {$_5$};
\draw[-stealth'] (0.5,0.4) to  [bend right] (-0.5,0.4);
\node[left] at (-0.5,0.4) {$_6$};
\end{scope}
\node[below] at (3,0) {$L$};
\end{tikzpicture}
\end{center}
 \caption{A standard plane beam element with its 6 degrees of freedom}
\label{fig:beamelement}
\end{figure}

The stiffness matrix $[K]_b$ of a beam element, including bending and axial effects, with respect to the nodal degrees of freedom shown in Figure \ref{fig:beamelement} can be conveniently expressed as the sum of axial and bending contributions, namely,
\begin{equation}
[K]_b=[K]_{axial}+[K]_{bend}
\end{equation}
with 
\begin{equation}
[K]_{axial}= EA_s \left[
\begin{matrix}
1/L&0&0&-1/L&0&0\\0&0&0&0&0&0\\0&0&0&0&0&0\\
-1/L&0&0&1/L&0&0\\0&0&0&0&0&0\\0&0&0&0&0&0
\end{matrix}
\right]
\end{equation}
and
\begin{equation}
[K]_{bend}= EI \left[
\begin{matrix}
0&0&0&0&0&0\\0&12/L^3&6/L^2&0&-12/L^3&6/L^2\\0&6/L^2&4/L&0&-6/L^2&2/L\\
0&0&0&0&0&0\\0&-12/L^3&-6/L^2&0&12/L^3&-6/L^2\\0&6/L^2&2/L&0&-6/L^2&4/L
\end{matrix}
\right]
\end{equation}
where $A_s$ is the cross-sectional area. For positive values of the constants involved, the ranks of these matrices are $1$ and $2$, respectively. Their sum has rank 3, indicating that there remain $6-3$ independent rigid-body motions in the plane. The lower ranks of the axial and bending components have a clear meaning, since they disregard either the bending or the axial resistance to deformation. We remark that, on physical grounds, these matrices must be positive semi-definite. In other words, they have no negative eigenvalues.

\subsection{The total stiffness matrix}

It is not difficult to obtain the stiffness matrix of the structure by assembling the stiffness matrices of the three bars, using the standard procedures of structural analysis. The entries in the blocks $[A]$ and $[B]$ can be split into axial and bending contributions, that is,
\begin{equation}
\begin{split}
[A]&=[A]_{axial}+[A]_{bend} \\ &=  EA_s \left[
\begin{matrix}
3/2d&0&0\\0&1/2d&0\\0&0&0
\end{matrix}
\right] + EI\left[
\begin{matrix}
6/d^3&0&0\\0&18/d^3&-6 \sqrt{3}/d^2\\0&-6\sqrt{3}/d^2&8/d
\end{matrix}
\right]
\end{split}
\end{equation}
and
\begin{equation}
\begin{split}
[B]&=[B]_{axial}+[B]_{bend} \\ &= EA_s \left[
\begin{matrix}
3/4d&\sqrt{3}/4d&0\\-\sqrt{3}/4d&-1/4d&0\\0&0&0
\end{matrix}
\right] + EI\left[
\begin{matrix}
-3/d^3&3\sqrt{3}/d^3&-3/d^2\\-3\sqrt{3}/d^3&9/d^3&-3 \sqrt{3}/d^2\\3/d^2&-3\sqrt{3}/d^2&2/d
\end{matrix}
\right]
\end{split}
\end{equation}
Correspondingly, the total structural stiffness matrix of the frame
can be split as
\begin{equation} \label{eq36}
\begin{tikzpicture}[scale=1.1]
\node at  (-2.5,1.5) {$[K]\;=[K]_{axial}+[K]_{bend}\;=$};
\draw[thick] (-0.1,0.1)--(-0.1,2.9);
\draw[thick] (-0.1,0.1)--(0,0.1);
\draw[thick] (-0.1,2.9)--(0,2.9);
\draw[thick] (3.1,0.1)--(3.1,2.9);
\draw[thick] (3,0.1)--(3.1,0.1);
\draw[thick] (3,2.9)--(3.1,2.9);
\foreach \x in {1,2}
{\draw (0,\x)--(3,\x);
\draw (\x,0.2)--(\x,2.8);}
\node at (0.5,2.5) {$A_{axial}$};
\node at (1.5,2.5) {$B_{axial}$};
\node at (0.5,1.5) {$B_{axial}^T$};
\node at (2.5,2.5) {$B_{axial}^T$};
\node at (1.5,1.5) {$A_{axial}$};
\node at (2.5,1.5) {$B_{axial}$};
\node at (0.5,0.5) {$B_{axial}$};
\node at (1.5,0.5) {$B_{axial}^T$};
\node at (2.5,0.5) {$A_{axial}$};
\begin{scope} [xshift=4cm]
\node at (-0.5,1.5) {$+$};
\draw[thick] (-0.1,0.1)--(-0.1,2.9);
\draw[thick] (-0.1,0.1)--(0,0.1);
\draw[thick] (-0.1,2.9)--(0,2.9);
\draw[thick] (3.1,0.1)--(3.1,2.9);
\draw[thick] (3,0.1)--(3.1,0.1);
\draw[thick] (3,2.9)--(3.1,2.9);
\foreach \x in {1,2}
{\draw (0,\x)--(3,\x);
\draw (\x,0.2)--(\x,2.8);}
\node at (0.5,2.5) {$A_{bend}$};
\node at (1.5,2.5) {$B_{bend}$};
\node at (0.5,1.5) {$B_{bend}^T$};
\node at (2.5,2.5) {$B_{bend}^T$};
\node at (1.5,1.5) {$A_{bend}$};
\node at (2.5,1.5) {$B_{bend}$};
\node at (0.5,0.5) {$B_{bend}$};
\node at (1.5,0.5) {$B_{bend}^T$};
\node at (2.5,0.5) {$A_{bend}$};
\end{scope}
\end{tikzpicture}
\end{equation}

Significantly, a homogeneous extension (that is a vector with components $\{1,0,0,1,0,0,1,0,0\}$) is an eigenvector of $[K]_{bend}$ corresponding to a vanishing eigenvalue. Physically, since a homogeneous expansion involves no bending of any of the beams, the total bending stiffness contributes no strain energy. The rank of $[K]_{bend}$ is, accordingly, 5. Conversely, all non-zero vectors of the form $\{0,0,a,0,0,b,0,0,c\}$ are eigenvectors with zero eigenvalues of $[K]_{axial}$, since no axial elongations take place. The rank of $[K]_{axial}$ is 3. We will presently consider the implications of these observations when dealing with self-similar fractals.

\section{Multi-scaled stiffness of a self-similar fractal}
\label{sec:gasket}

\subsection{Geometrical self-similarity and its consequences}

\begin{figure}[H]
\begin{center}

\def\trianglewidth{2.5cm}%
\pgfdeclarelindenmayersystem{Sierpinski triangle}{
    \symbol{X}{\pgflsystemdrawforward}
    \symbol{Y}{\pgflsystemdrawforward}
    \rule{X -> X-Y+X+Y-X}
    \rule{Y -> YY}
}%
\foreach \level in {0,...,4}{%
\tikzset{
    l-system={step=\trianglewidth/(2^\level), order=\level, angle=-120}
}%
\begin{tikzpicture}
    \fill [black] (0,0) -- ++(0:\trianglewidth) -- ++(120:\trianglewidth) -- cycle;
    \draw [draw=none] (0,0) l-system
    [l-system={Sierpinski triangle, axiom=X},fill=white];
\end{tikzpicture}
}%

\end{center}
 \caption{Generation process of a Sierpi\'nski triangle}
\label{fig:gasket}
\end{figure}
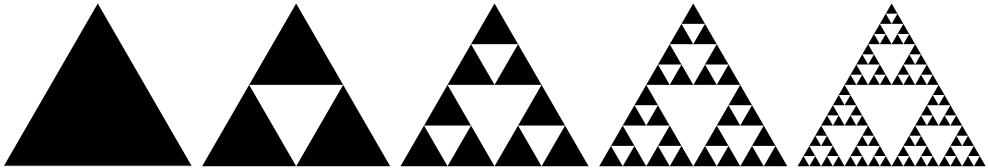

The Sierpi\'nski triangle (also known as the Sierpi\'nski gasket) is a subset of the Euclidean plane obtained by recursively removing from a solid equilateral triangle its central (half-sized) triangular portion and repeating this procedure ad infinitum in each of the remaining solid triangles. Figure \ref{fig:gasket} illustrates the first few steps of the construction.

The considerations of symmetry and equilibrium, carefully detailed in Section \ref{sec:frame} for the case of a triangular frame, apply equally well to a material Sierpi\'nski gasket or, for that matter, to any figure enjoying the geometrical and material symmetries of the equilateral triangle under the general assumption of linearity between causes (in-plane displacements and rotations) and effects (in-plane forces and couples). We have found that, by virtue of those considerations alone, the stiffness matrix relative to the 9 specified nodal degrees of freedom is determined by 5 independent entries. Moreover, the general block structure of the stiffness matrix is expressed in terms of two $3 \times 3$ blocks of the form given by Equations (\ref{eq25}) and (\ref{eq26}).

We remark that, because of the inhomogeneity of the physical units of rotations and displacements, the entries in the stiffness matrix $[K]$ are also unit-wise inhomogeneous. We have, in fact, already observed in Section \ref{sec:simplest} that the stiffness matrix of a standard plane beam of length $L$, contains terms proportional to $1/L, 1/L^2$, and $1/L^3$, and we found it convenient to introduce non-dimensional stiffness coefficients. To facilitate the treatment, therefore, we define non-dimensional entries $\alpha_1,\alpha_2,\alpha_3,\alpha_4,\beta_1,\beta_2,\beta_3,\beta_4,\beta_5$ and $\beta_6$ as
\begin{equation} \label{eq:nondim}
\begin{split}
\alpha_1=\frac{a_1}{a_1}=1\;\;\;\;\;\;\alpha_2=\frac{a_2}{a_1}\;\;\;&\;\;\alpha_3=\frac{a_3}{a_1 d^2}\;\;\;\;\;\;\alpha_4=\frac{a_4}{a_1 d} \\ \\
\beta_1=\frac{b_1}{a_1}\;\;\;\;\;\;\beta_2=\frac{b_2}{a_1}\;\;\;\;\;\;\beta_3=\frac{b_3}{a_1 d}\;\;\;&\;\;\beta_4=\frac{b_4}{a_1}\;\;\;\;\;\;\beta_5=\frac{b_5}{a_1 d}\;\;\;\;\;\;\beta_6=\frac{b_6}{a_1 d^2}
\end{split}
\end{equation}
where $d$ denotes the side of the triangle. In terms of these variables the equilibrium relations (\ref{eq:equil}) can be rewritten as
\begin{equation} \label{eq:equilnondim}
\begin{split}
\beta_2&=\frac{\sqrt{3}}{3} (\alpha_1-\beta_1) \\
\beta_3&=\frac{\sqrt{3}}{2}\alpha_4 -\left(\frac{1}{3}(\alpha_1-\beta_1) -\frac{1}{2} \alpha_2\right)\\
\beta_4&=a_2-(a_1-\beta_1)\\
\beta_5&=-\frac{1}{2}\alpha_4+\sqrt{3}\left(\frac{1}{3}(\alpha_1-\beta_1)-\frac{1}{2}\alpha_2 \right)\\
\beta_6&=-\frac{1}{2}\alpha_3-\left(\frac{1}{3}(\alpha_1-\beta_1)-\frac{1}{2}\alpha_2\right)
\end{split}
\end{equation}
Similarly, the basic matrix blocks, $[A]$ and $[B]$, can be expressed as
\begin{equation} \label{eq:blocksab}
[A]=a_1\,[\Gamma][ \alpha][\Gamma]\;\;\;\;\;\;\;\;\;\;[B]=a_1\,[\Gamma][\beta][\Gamma]
\end{equation}
where
\begin{equation} \label{eq:blocksabnondim}
[\alpha]=\left[
\begin{matrix}
\alpha_1&0&0 \\0&\alpha_2&\alpha_4\\ 0&a_4&\alpha_3
\end{matrix}
\right]
\;\;\;\;\;\;\;\;
[\beta]=\left[
\begin{matrix}
\beta_1&\beta_2&\beta_3 \\-\beta_2&\beta_4&\beta_5\\ -\beta_3&\beta_5&\beta_6
\end{matrix}
\right]
\;\;\;\;\;\;\;\;
[\Gamma]=\left[
\begin{matrix}
1&0&0\\0&1&0\\ 0&0&d
\end{matrix}
\right]
\end{equation}

\subsection{Enlargement and condensation}

\begin{figure} [H]
\begin{center}
\begin{tikzpicture}

\draw[thick, fill=gray!40] (0,0)--(4,0)--(2,{2*sqrt(3)})--cycle;
\draw[thick, fill=white] (2,0)--(3,{sqrt(3)})--(1,{sqrt(3)})--cycle;
\node[above] at (2,{2*sqrt(3)}) {$I$};
\node[left] at (0,0) {$J$};
\node[right] at (4,0) {$K$};
\node[right] at (3,{sqrt(3)}) {$_{I'}$};
\node[left] at (1,{sqrt(3)}) {$_{J'}$};
\node[below] at (2,0) {$_{K'}$};
\draw[|-|] (0,-0.5)--(4,-0.5);
\node[below] at (2,-0.5) {$2d$};
\end{tikzpicture}
\end{center}
 \caption{A Sierpi\'nski gasket as an assembly of three smaller copies}
\label{fig:sierpinski1}
\end{figure}
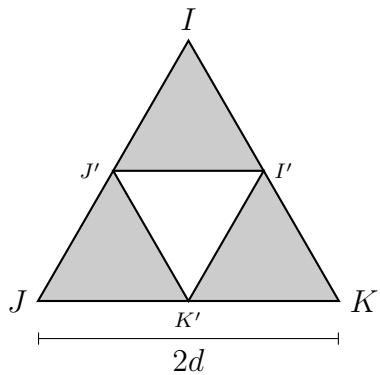

Since a Sierpi\'nski triangle consists of an assembly of 3 reduced-by-half copies of itself, and since the 3 smaller copies are joined only at their vertices, as shown in Figure \ref{fig:sierpinski1}, the stiffness matrix of the larger triangle (with side $2d$) with respect to its nodal degrees of freedom at $I,J,K$ can be obtained by a process of static condensation, as already illustrated for the simpler case of Section \ref{sec:condense}, resulting in the elimination of the degrees of freedom at nodes $I',J',K'$.

The result of this assembly of 3 identical copies of the smaller triangle and the eventual condensation of the intermediate degrees of freedom is expressed as a definite mathematical relation between the condensed stiffness matrix of the larger triangle and the stiffness matrix of the smaller one. If the smaller triangular structures, even if identical to each other, were not self-similar, there would be no basis to justify a cognitive leap of faith of the kind contemplated in our simplest example. If, on the other hand, the smaller constituents are  geometrically self-similar, one would be precipitous in cavalierly assuming that the principle of stiffness self-similarity between the larger and the smaller entities can be applied. Indeed, from the (non self-similar) triangular frame of Section \ref{sec:frame}, we have learned that the stiffness contribution associated with bending (controlled in that example by the moment of inertia) will in general undergo a scaling upon geometric enlargement different from the counterpart contributed by axial stretch (controlled by the cross-sectional area). 

For a structure that includes different mechanisms contributing to the structural stiffness, one may venture to propose the following
\newline

{\bf Generalized principle of stiffness self-similarity}: Each additive contribution to the total stiffness matrix will separately abide by the principle of stiffness self-similarity, with its own stiffness-scaling factor.
\newline

At first sight, this principle appears to be of a problematic nature. Indeed, one may ask: are these different mechanisms identifiable a priori? If so, are the individual stiffness-scaling factors (or, at least, their ratios) available? These questions will be addressed in the next section.

\subsection{The fine points}

Our objective is to determine the generic stiffness matrix $[K]$ of a material Sierpin\'ski gasket of side $d$ deforming in its plane and incorporating as nodal unknowns the two displacement components and the in-plane rotation of each node. The inclusion of the latter is motivated by the aim to produce a viable shell-like fractal triangle that can be used to fit a polyhedral tiling to a surface. The transverse bending modes (with two degrees of freedom of rotation and one of transversal displacement) have been independently treated in \cite{plate}. The virtue of the present treatment is that it furnishes not only the missing (in-plane) displacement components but also the so-called drilling mode.

The desired $6 \times 6$ stiffness matrix $[K]$ is of the form (\ref{eq:blocksab}). It is completely determined by specifying the five entries $a_1, a_2, a_3, a_4$ and $b_1$. The remaining 5 entries, namely $b_2,b_3,b_4, b_5$ and $b_6$, are obtained from Equations (\ref{eq:equil}) or their non-dimensional counterpart (\ref{eq:equilnondim}). Assuming that the values of $a_1, a_2, a_3, a_4$ and $b_1$ have been given, we proceed to assemble 3 copies of the gasket to produce a new gasket of side $2d$, as in Figure \ref{fig:sierpinski1}. This structure, involving 6 nodes and 18 degrees of freedom, will be subjected to a process of static condensation to eliminate the intermediate (mid-side) degrees of freedom.

To carry out the aforementioned condensation it is convenient to express the nodal degrees of freedom in terms of components in a global coordinate system $x,y,z$, such as the one used in Figure \ref{fig:frame}, where the $z$-axis runs perpendicularly to the page and points towards the reader. The components of vectors in the new coordinate system are obtained by multiplying the old components by the matrix $R_\phi$ given by
\begin{equation}
[R_\phi]=\left[
\begin{matrix}
\cos{\phi}&\sin{\phi}&0\\-\sin{\phi}&\cos{\phi}&0\\0&0&1
\end{matrix}
\right]
\end{equation}
where $\phi$ is the counter-clockwise angle of rotation about $z$ needed to bring the old coordinate axes into the new ones.
The $9 \times 9$ rotation matrix $[R]$ to be applied to transform the stiffness $[K]$ to the global coordinates according to $[K']=[R]^T[K] [R]$ is obtained in block form as
\begin{equation}
[R]=\left[
\begin{matrix}
R_{270^o}&0&0\\0&R_{150^o}&0\\ 0&0&R_{30^o}
\end{matrix}
\right]
\end{equation}
Assembling the stiffness matrix corresponding to the 18-degree-of-freedom structure depicted in Figure \ref{fig:sierpinski1} we obtain an $18 \times 18$ stiffness matrix which can be partitioned into four $9 \times 9$ blocks as
\begin{equation}
[{\mathbb K}]=\left[
\begin{matrix}
L&M\\M^T&N
\end{matrix}
\right]
\end{equation}
where the symmetric blocks $L$ and $N$ correspond, respectively, to the stiffness coefficients associated with the degrees of freedom of nodes $I,J,K$ and $I',J',K'$, while the blocks $M$ and $M^T$ are the cross interactions between both sets of nodes. The condensed $9 \times 9$ stiffness matrix is obtained as
\begin{equation}
[K]'_c=[L]-[M][N]^{-1}[M]^T.
\end{equation}
We remark that this matrix can be brought back to the original degrees of freedom according to the inverse transformation $[\hat K]=[R][K]'_c[R]^[R]^T$. At the end of this process, the condensed matrix $[K]_c$ must necessarily be of the form given in Equation (\ref{eq:blocksab}) and its entries must satisfy the equilibrium conditions.  

All the operations just described are easily programmable, for example, into a Mathematica\textsuperscript{\textregistered} module. As a result, we have two $9 \times 9$ matrices, $[K]$ and $[\hat K]$ corresponding, respectively, to the original gasket and to its enlarged version (of side $2d$). In effect, the program module can be regarded as the evaluation of 5 functions for the entries $\hat a_1,\hat a_2, \hat a_3, \hat a_4$ and $\hat b_1$ in terms of $a_1, a_2, a_3, a_4$ and $b_1$.
If the system were to abide by the ordinary principle of stiffness self-similarity, we would simply have to solve a system of 4 simultaneous equations establishing the equality of the non-dimensional versions of the respective coefficients. A solution of this system is best found by a numerical procedure such as the method of Newton-Raphson, again not a difficult task to program. We would find not only the values of these coefficients but also the stiffness scaling factor $\hat a_1/a_1$.

If the system is endowed with multi-scaled stiffness, what would a solution of the above mentioned algebraic system represent? It can only represent the particular case of one of the stiffness mechanisms at play! No non-trivial combination of these mechanisms can abide by a single common stiffness scaling. By analogy with the case of the triangular frame discussed in Section \ref{sec:frame}, we would have found either the mode corresponding to a vanishing moment of inertia or, alternatively, to the vanishing of the cross sectional area. Moreover, on physical grounds, we would expect that the stiffness matrix corresponding to one of the modes be of rank 3 and the matrix corresponding to the other mode be of rank 5. In other words, the solution algorithm itself answers all the pertinent questions, including the determination of the constituent stiffness matrices and the corresponding scalings.

Carrying out the steps just described, two (numerical) solutions were found, to wit,
\begin{enumerate}
\item {\bf The axial mode}:
\begin{equation}\label{eq45}
\begin{split}
&[\alpha]=\left[
\begin{matrix}
1&0&0\\0&0.333333&0\\0&0&0
\end{matrix}
\right], \\  \\ &[\beta]=\left[
\begin{matrix}
0.5&0.2886725&0\\-0.2886725&-0.166666&0\\0&0&0
\end{matrix}
\right], \\ \\
&\;\, \frac{\hat a_1}{a_1}=\;0.5
\end{split}
\end{equation}
\item {\bf The bending mode} 
\begin{equation}\label{eq46}
\begin{split}
&[\alpha]=\left[
\begin{matrix}
1&0&0\\0&1.45714&-0.593846\\0&-0.593846&0.376471
\end{matrix}
\right], \\  \\ &[\beta]=\left[
\begin{matrix}
-0.5&0.866025&-0.285714\\-0.866025&-0.0428571&-0.0989743\\0.285714&-0.0989743&0.0403361
\end{matrix}
\right], \\ \\
&\;\, \frac{\hat a_1}{a_1}=\;0.15
\end{split}
\end{equation}
\end{enumerate}

The stiffness ratio $\kappa=\hat a_1/a_1$ corresponds to the doubling of the geometric ratio $\rho$. We surmise, accordingly, that the dependence of $\kappa$ on $\rho$ for our fractal triangle with given elastic material properties abides by the formula
\begin{equation}
\kappa(\rho)=\kappa^{\;ln_{_2} \rho}
\end{equation}

\begin{rem} {\rm The entries in the axial and bending non-dimensional matrices were obtained by strictly implementing the outlined numerical procedure. It is comforting to be able to check that the non-zero entries in the axial matrices $[\alpha]$ and $[\beta]$ are consistent with the exact values derived analytically for the Sierpi\'nski gasket in the absence of rotational degrees of f
freedom. As listed in \cite{paper1}, these $2 \times 2$ matrices are
\begin{equation} 
[\alpha]_o=\left[
\begin{matrix}
1&0 \\ 0&1/3 
\end{matrix}
\right] \;\;\;\;\;\;\;\;\;\;\;\;\;\; [\beta]_o = \left[
\begin{matrix}
1/2&\sqrt{3}/6\\-\sqrt{3}/6&-1/6
\end{matrix}
\right]
\end{equation}
}
\end{rem}

\begin{rem} {\rm The numerical solutions (\ref{eq45}) and (\ref{eq46}) listed above were obtained by an impartial `brute force technique', whereby all solutions are sought that satisfy the equality of the non-dimensional matrices $[\alpha]$ and $[\beta]$ of the original and condensed structures. The starting values of the iterative procedures were chosen at random. If, however, an a priori knowledge of some features of the underlying stiffness mechanisms is available, the numerical procedure can be significantly improved as it is targeted to one of these specific mechanisms. In our case, for instance, if one recognizes on physical grounds that the bending mechanism will turn out to be indifferent to a homogeneous expansion of the gasket, one can predict that the vector of degrees of freedom with entries $\{1,0,0,1,0,0,1,0,0\}$ will be an eigenvector corresponding to a zero eigenvalue of the bending part of the stiffness matrix. In other words, the restriction $a_1+2b_1=0$ can be imposed ab initio. The problem is then downgraded to the application of the ordinary principle of stiffness self-similarity. The convergence of the numerical procedure can be thus dramatically accelerated, as we were able to verify independently from the previous calculations.}
\end{rem}

\section{Final thoughts}
\label{sec:final}
In closing, it is fair to make some remarks pertaining to the features and also to the limitations of our formulation. A salient feature of the approach adopted is its simplicity and its obvious connection to the standard theory of classical structural mechanics, except for the added exploitation of self-similarity. On the other hand, it does not suggest any clues as to how it may be extended to general (not necessarily self-similar) fractals, nor does it offer an avenue to a generalization for geometrical and/or material non-linearity. The displacement of any internal point in the fractal can be obtained by an exact and relatively simple process of recursion. Nevertheless,  it is of the very nature of fractal mechanics that the collection of all such displacements does not constitute a smooth field, nor is it amenable to approximation by smooth interpolation functions. These and other considerations being made, if an engineer is interested in creating a shell consisting of a surface tiled with triangles connected at their nodes, and if the features of the material and the technique of its deposition (spraying, say) are such to suggest the emergence of fractal properties, all that is needed to obtain the nodal displacements and rotations is the assembly of the total structural stiffness matrix. Since the concept of stress is not trivially available for fractals, criteria of failure may be based on stored energy per unit volume considerations.

\end{document}